# Surface quality improvement of porous thin films suitable for nanoindentation

Zhangwei Chen*, Xin Wang, Finn Giuliani, Alan Atkinson

Department of Materials, Imperial College London SW7 2BP, UK
*Corresponding author: chen@ic.ac.uk (Z. Chen)

**Abstract**

The reliability of perovskite material $La_{0.6}Sr_{0.4}Co_{0.2}Fe_{0.8}O_{3-\delta}$(LSCF) to be used as cathode parts in solid oxide fuel cells (SOFCs)also relies on its mechanical properties. Adequate surface conditions (i.e. flat and crack-free) are desired when the as-sintered porous thin films are subjected to nanoindentation for mechanical property determination. In this study, extensive cracks and considerable surface roughness were found in the LSCF films after sintering at high temperatures. This would significantly scatter the nanoindentation data and result in unreliable measurements. Various attempts including the comparison of film deposition methods, drying and sintering processes, and reformulating the ink were made to improve the surface quality. Results revealed little dependence of cracking and surface roughness on deposition methods, drying or sintering processes. It was found that the critical factor for obtaining crack-free and smooth LSCF films was the ability of the ink to be self-levelling in the earlier wet state. Reproducible nanoindentation measurements were obtained for the films with improved surface quality.

## 1. Introduction

Thanks to its excellent mixed electronic-ionic conductivity[1], perovskite material $La_{0.6}Sr_{0.4}Co_{0.2}Fe_{0.8}O_{3-\delta}$(LSCF) has been considered as a promising candidate to be used as cathode parts in intermediate temperature solid oxide fuel cells (IT-SOFCs)[2].Besides, an adequate porosity of approximately 30-40 vol% and a compatible thermal expansion coefficient matching with that of electrolytes are also desired along with a low cost and simple fabrication process[3, 4]. Currently most LSCF cathode films, often thin and porous, are fabricated by depositing (e.g. screen-printing)a few layers of commercial inks onto dense sintered electrolyte substrates, such as gadolinium doped ceria ($Ce_{0.9}Gd_{0.1}O_{2-\delta}$, CGO), followed by high temperature sintering[3]. The knowledge of the mechanical properties of such cathode films is a prerequisite for the reliability and durability of the parts to be assessed for their application in rigorous environments where mechanical stresses arise.

Nanoindentation is an advantageous technique with high spatial and depth resolution that can be used to measure the mechanical properties (e.g. elastic modulus and hardness) of very small volumes of material, such as thin films of a few micrometers thickness[5, 6].There is currently no report on response of nanoindentation on porous thin LSCF films. In addition, in order that nanoindentation tests be reproducible and thus the results be reliable, cracking and surface asperities (i.e. agglomerates) in the as-sintered test films must be avoided or minimized to an acceptable level, which means the size of such defects should be much less comparable to the indented feature such as indentation depth, which in the current study wasless than 2 μm. However, as can be seen later in this paper, such defects were found constantly in the sintered LSCF films, which severely scattered the nanoindentation response and thus the resulting measurements were considered invalid.

Although such defects in the as-sintered LSCF films were also observed in some other studies, the problem has not yet been considered critical in many researches which essentially involved the study of electrochemical performance rather than mechanical properties of the SOFC component, as discussed later in Section 3. Due to the nature of the fabrication process and the application in SOFCs, the LSCF cathode films are difficult to be surface treated to obtain relatively flat surfaces, and the cracking remains a problem. As a result, the relevant technological challenges could be related to the fabrication of thin, porous, flat and crack-free LSCF cathode films.

To mitigate the cracking and surface asperity problems in LSCF cathode films, in the current study efforts were made in various aspects which involved the comparison of deposition methods, the improvement of drying and sintering processes, and of the ink formulation. Results revealed that a sufficiently low viscosity of the ink could result in crack-free LSCF films with flatter surfaces and a more homogeneous microstructure, while the alternative deposition method and the optimized thermal treatments applied helped little to solve the problems. In other words, the surface quality of the films was significantly improved by using a reformulated less viscous ink which ensured more ability to self-level in the early wet state upon deposition. By using these non-defective films, effective nanoindentation measurements were achieved with high reproducibility and hence reliable results were obtained.

## 2. Methods and Materials

Fully dense CGO pellet samples of approximately 25 mm in diameter and 3 mm thickwere prepared to be used as electrolyte substrates for LSCF film deposition by uniaxially pressing the powder (NexTech Materials, USA) at 150 MPa pressure for 10s. They were then isostatically pressed at 200 MPa for further compaction for 30s, followed by sintering at1400 °C in air with a heating and cooling rate of 5 °C/min and a holding time of 4 hours. The as-sintered dense substrates were polished on one face by successively using 15, 9, 6, 4 and 1 micron diamond suspensions, to generate mirror smooth and flat surfaces.

An LSCF ink from ESL, UK was deposited first on the fully dense CGO substrates using either screen printing with 250 mesh screen and 2.5 mm gap or tape casting with a perimeter mask of 40 μm height. The ready-made films were then oven-dried at 100 °C for 12 hours, followed by sintering in air at 1000 °C for 4 hours with a heating and cooling rate of 5 ºC/min, according to the initial sintering program.

Reformulated inks were prepared respectively by diluting the as-supplied LSCF ink 1:1and 1:2by volume with terpineol (Sigma, UK) and then ball-milled for 12 hours to reach homogeneity. These two inks are noted as 1:1 ink and 1:2 ink in this study. The viscosity of the inks was measured by controlled-stress rheometer (CVO100D, Malvern Instruments Ltd, UK) at room temperature under shear rate of $95s^{-1}$. The same deposition processes (i.e. screen printing and tape casting) were carried out for the reformulated inks to produce films for subsequent comparison experiments.

Thermogravimetric analysis (TGA) of the LSCF ink was performed in air in order to determine the thermal decomposition range. Dilatometry analysis on uniaxial-pressed green bodies of LSCF powders were carried out using a push-rod dilatometer DIL-402C (NETZSCH, Germany) to investigate the sintering activity and shrinkage of the LSCF material. The experiment temperature range was set between room temperature and 1250 °C, at four constant heating rates of 3, 5, 10 and 20 °C/min.

The microstructures of the dried and as-sintered films were studied using scanning electron microscopy (SEM) JEOL 5610 (JEOL, Japan), and the cross-sectional surfaces were preparedby fracture or focused ion beam (FIB) slice and view using Helios Nanolab (FEI, USA).The width of crack opening was also determined under SEM. The average surface roughness ($R_a$) of the sintered LSCF films was measured using an optical interference surface profiler OMP-0360G (Zygo, USA).

Nanoindentation tests were performed under load control mode up to 500 mNon a NanoTest platform (Micromaterials, UK) at room temperature using a spherical diamond tip of 50 μmin diameter, with 1 mN/s loading and unloading rate. Holding times of 2 seconds and 20 seconds were applied at peak load and at the end of unloading, respectively, to check for the creep effect and thermal drift. For each sample, a set of nanoindentation tests on at least 20 different locations on the surface were performed to obtain more representative average results. Before each test, calibration was made on a standard silica sample to establish the system frame compliance. The resulting nanoindentation response (i.e. load versus indentation depth curves) and the elastic modulus which was determined by applying Olive-Pharr method[5] based on the curves, were then used for the assessment of the reliability and reproducibility of the nanoindentation tests on the films.

### 3. Results and Discussion

#### 3.1 Defects in the as-sintered LSCF films

In the preliminary study, severe cracking networks and a large number of surface asperities (i.e. agglomerates) were generated in the sintered films deposited by screen printing using the as-received ink. The surface and cross-sectional microstructures of a film after sintering at 1000 °C are shown in Fig. 1. The width of crack openings and the size of the agglomerated islands were measured to be 1-4 μm and 20-40 μm, respectively. It should be noted, however, that there was a good adhesion between the film and the substrate. Fig. 2 shows a magnified cross-section sliced by FIB, revealing vertical cracks penetrating through the film thickness and a dense agglomerate present under the film surface.

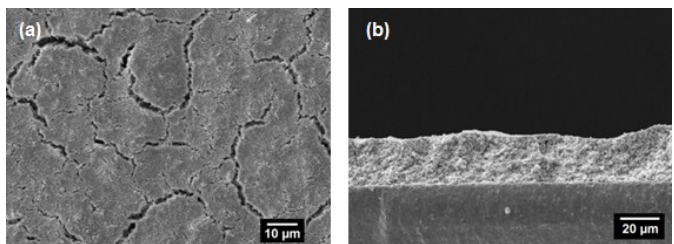

*Fig. 1. SEM micrographs of a LSCF film after sintering at 1000 °C: (a) top surface view of crack networks and agglomerates; (b) fracture cross-sectional view of the poor surface smoothness*

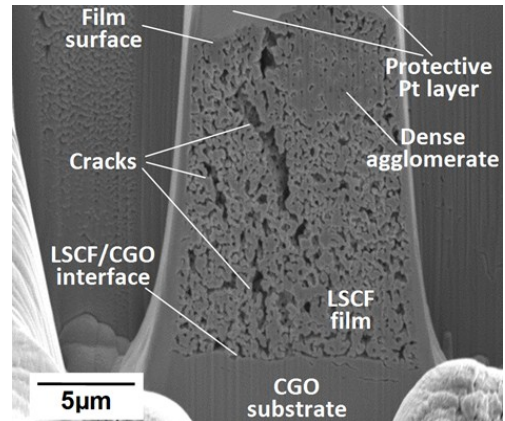

*Fig. 2. FIB sliced cross-section showing cracks penetrating through the film and a large asperity near the surface of a film sintered at 900 °C for 4h*

The fabrication of cathode components of SOFCs has been reported using various methods, such as screen printing, tape casting, dip casting, chemical vapor deposition, atmospheric plasma spray, colloidal spray deposition, pulsed-laser deposition, sputtering and painting, as reviewed by Fergus et al. [3] and Taroco et al.[7]. In Table 1 we summarize the techniques reported to be used for the deposition of LSCF films on CGO substrates [8-13].

*Table 1. Summary of observations from the literature on the deposition of LSCF films on CGO substrates.*

| Reference | Deposition method | Heat treatment conditions | Film thickness (μm) | Crack opening width (μm) | Surface asperities |
|---|---|---|---|---|---|
| Lee et al. [8] | Screen printing | Sintering at 1000°C for 0.5h | < 40 | < 30 | Yes |
| Baque et al. [9][a] | Dip coating | Dried at 130 °C; Sintering at 900 °C for 6 h | < 28 | < 20 | Yes |
| Marinha et al. [10] | Electrostatic spray deposition (ESD) | Substrate temperature 250-450 °C; Sintering at 900 °C for 2h | < 25 | < 5 | Yes |
| Hsu et al. [11] | Electrostatic-assisted ultrasonic spray pyrolysis (EAUSP) | Deposition temperature 350 °C; Sintering at 1000°C for 2 h | -- | < 2 | Yes |
| Santillan et al. [12] | Electrophoretic deposition (EPD) | Sintering at 950°C in air for 2 h | < 25 | No detectable crack | Yes |
| Wang et al. [13] | In-situ solid-state reaction sintering | Sintering at 1000 °C in air for 6 h | -- | No detectable crack | Yes |

[a]$La_{0.4}Sr_{0.6}Co_{0.8}Fe_{0.2}O_{3-\delta}$ was used *as film material in this case.*

From the literature listed above, it can be concluded that most LSCF films possessed more or less cracking and surface asperities after final heat treatments, irrespective of the deposition method applied. Although no detectable cracking was present in the films for [12] and [13], the surface asperities remain a great concern when it comes to undertaking nanoindentation tests.

Adjusting working parameters may to some extent reduce film cracking, but these deposition methods still struggle with preparing crack-free and flat LSCF films containing highly porous structures. This implies that the existence of cracking is prevalent in many LSCF films irrespective of their deposition techniques. Therefore, the challenge remains to be the fabrication of crack-free and flat LSCF films. Although many have observed such defects in their LSCF films, they often attributed them to the densification and shrinkage of films during sintering, or the TEC mismatch between films and substrates during cooling, without any further experimental validation [8, 11], Although these defects might not be of major importance for studies on electrochemical performance, they are, however, crucial for the present study, as their presence in the films would certainly cause errors in the nanoindentation response. Furthermore they are likely to induce damage to cells in actual application.

### 3.2 Nanoindentation response on defective LSCF films

Nanoindentation on the film shown in Fig.1 was conducted and the response curves, namely the load vs. indentation depth curves, generated are shown in Fig.3.

The extremely irregular shapes and wide variability of the response curves seen in the figure were generated by the indenter tip touching defective locations on the films during nanoindentation tests. The presence of cracks and poor surface flatness due to large asperities induced significant variability in the measured elastic modulus values. In the current case, in addition to the extensive cracks, it was found that the average surface roughness $R_a$ for the defective film was measured to be 1.86 µm, which was very close to the indentation depth of approximately 2 µm, resulting in an enormous relative error as much as 63.2% for the apparent elastic modulus (34±21GPa), as shown in Fig. 3(b).

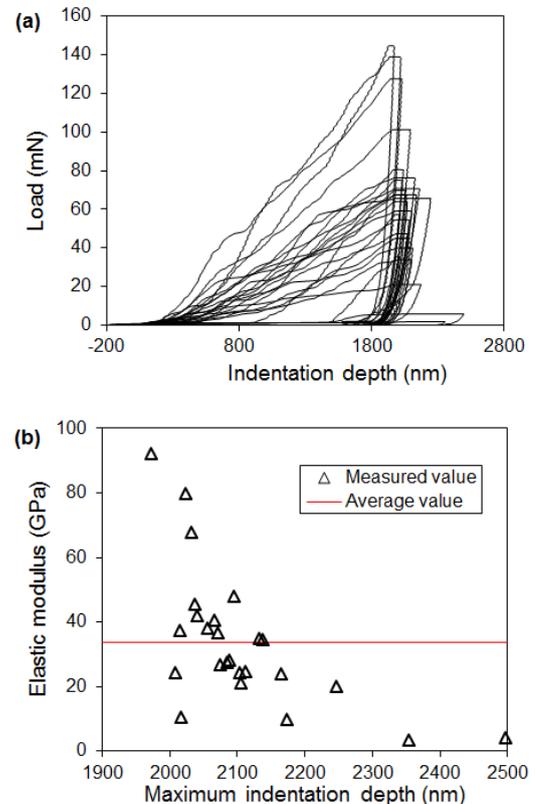

*Fig. 3. Nanoindentation load vs. indentation depth curves (a) and the corresponding elastic modulus values calculated (b) for the defective films similar to those in Figures 1 and 2.*

Examples of three typical nanoindentation load-depth curves collected in this study are plotted in Fig. 4, which demonstrate the existence of three different contact processes of the indenter tip with the sample surface and are related to the characteristic surface features described earlier.

Curve-1 is typical of a relatively smooth and un-cracked surface location, it being less influenced by these defects. However, there is still an irregular response during loading (such as the "pop-in" events, as indicated by arrows) which is attributed to the fact that the LSCF film was not smooth enough. In contrast, curve-2 and curve-3 show extremely unreliable responses. Curve-2 exhibits a deep "pop-in" attributable to the cracks in the contact location. The existence of the initial negative depth in curve-3 indicates that during loading an asperity on the indent surface was first touched by the tip.

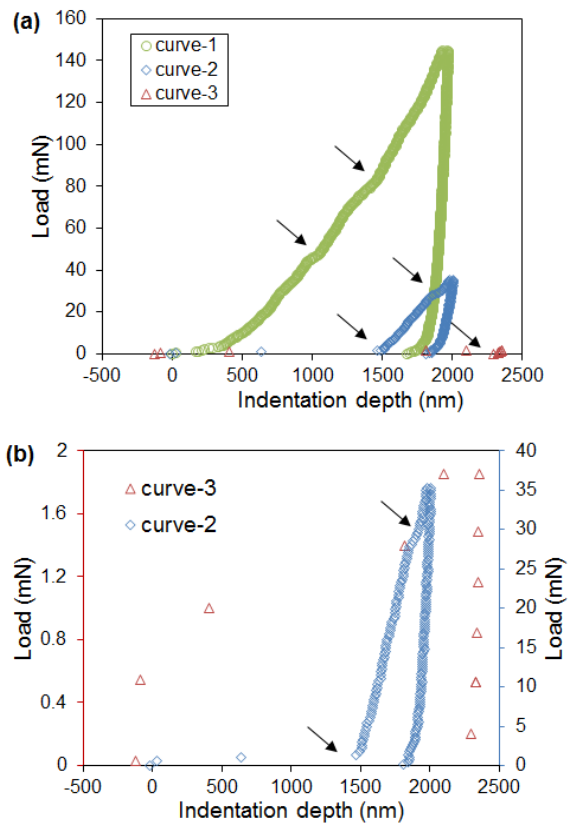

*Fig. 4. Examples of load-depth curves for nanoindentation using spherical indenter tip corresponding to likely features at the indent locations. (b) plots curve-2 and curve-3 on an expanded load scale.*

For a spherical indenter tip, the characteristic surface roughness of the test sample must be negligible compared to the maximum indentation depth. While for a Berkovich or other types of sharp tip, the indentation results are much more sensitive to the surface quality and porous microstructures. This was also the reason that in this work a spherical indenter tip was used for nanoindentation, particularly on porous LSCF films of which the surface was hard to polish.

**3.3 Improvement of LSCF film surface quality**
Given the analysis above, it was imperative that the cracking and surface asperities be reduced for the properties of the LSCF films to be reliably and reproducibly measured using nanoindentation. As the aim of this study, the improvement of the film surface quality required the investigation of the possible influencing factors of the defect formation, such as the TEC mismatch between LSCF films and CGO substrates, the drying, sintering and cooling processes applied, the deposition methods used as well as the formulation of the ink, which will be discussed below in more detail.

**3.3.1 TEC mismatch between LSCF films and CGO substrates**
One reason that LSCF is a promising cathode material is because its TEC ($15.3 \times 10^{-6}$ K$^{-1}$, 100-600 °C) [2] is close to that of CGO ($13.5 \times 10^{-6}$ K$^{-1}$) [14], which means there exists only a small thermal expansion mismatch. Nevertheless, even this small difference would induce a tensile thermal stress which would facilitate crack formation in the film during cooling after sintering. To avoid this possible effect caused by TEC mismatch, some dense substrates were prepared with the same material (i.e. LSCF) as the films by sintering at 1200°C for 4 hours with a heating rate of 300°C/h. The LSCF film was applied by screen printing the as-received ink and the resulting sample was then sintered under the same condition as the previous samples on CGO substrates. This would certainly exclude the crack formation caused by the TEC mismatch between films and substrates. From Fig.5, which shows SEM microstructures of the sintered LSCF film, it can be seen that the film surface was full of asperities.

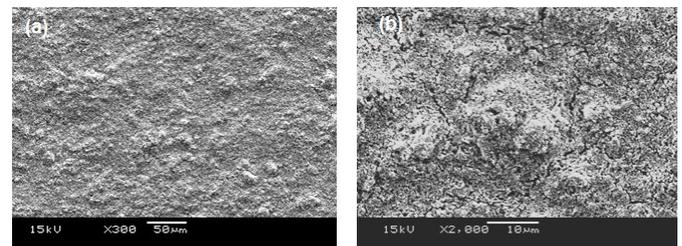

*Fig. 5. SEM images of sintered LSCF films screen printed on LSCF substrates. Images (a) and (b) are shown at different magnifications*

Again, a cracking network was observed on the surface with crack openings of 1.5 μm wide. Hence little dependence was shown of the cracking on the thermal expansion coefficient difference between LSCF and substrate materials since cracking still occurred for LSCF films on LSCF substrates, which involved zero TEC difference.

### 3.3.2 Drying, sintering and cooling processes
Cracking could also be attributed to the shrinkage of the films themselves during the drying and constrained sintering processes. Therefore, an investigation of the influence of thermal treatments on defect formation was carried out.

### 3.3.2.1 TG-DSC analysis
Thermogravimetric analysis (TGA) of the commercial LSCF ink was performed in air atmosphere in order to determine the thermal decomposition range of the organic constituents. Fig. 6 shows the TG-DSC curves of the ink at a heating rate of 10 °C/min.

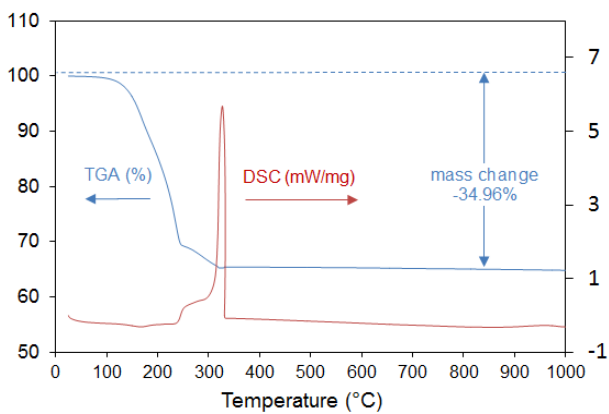

*Fig. 6. TG and DSC curves of the as-received LSCF ink in air atmosphere.*

As can be seen from the TG curve, the thermal decomposition of the ink underwent three steps: the volatile organic solvent started volatilization at around 100 °C, followed by oxidation/decomposition of a majority of organic content when the temperature increased to 220 °C, resulting in a major mass loss.The organic residues continued to be burnt-out until the temperature reached 320 °C, after which the weight remained almost constant.(The oxygen desorption from the LSCF perovskite lattice with increasing temperature[15] was too small to be observable on this scale.) The decomposition completed with a total mass loss of 34.96%.

### 3.3.2.2 Dilatometry
Dilatometry measurements on uniaxial-pressed green bodies of LSCF powder were performed to investigate the sintering activity and shrinkage of the LSCF powder. The heating dilatometric curves are shown in Fig. 7 at four constant heating rates: i.e. 3, 5, 10, 20 °C/min from room temperature to 1250 °C.

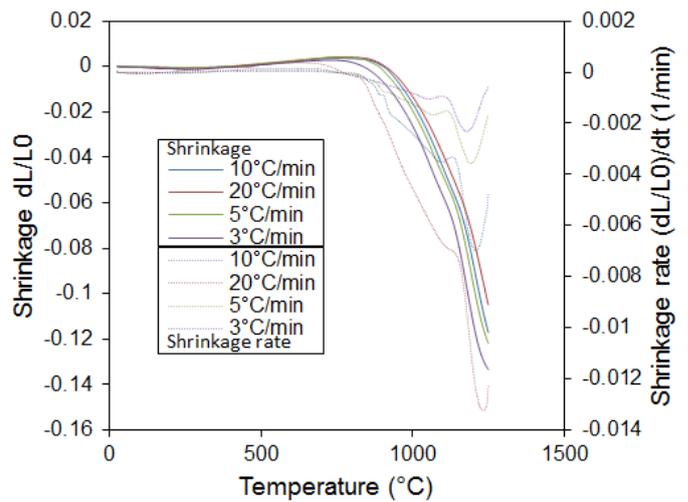

*Fig. 7. Shrinkage and shrinkage rate vs. temperature during constant heating rate sintering of LSCF at 3, 5, 10, 20 °C/min heating rates*

No significant dimensional changes were observed up to 500 °C, after which thermal expansion was observable until noticeable shrinkage phenomena began at around 800 °C. The densification process continued steeply up to 1250 °C. The final shrinkage ranged from 10% at a 20 °C/min heating rate to 14% at a 3 °C/min heating rate, suggesting that smaller heating rates allowed more sufficient time for shrinkage which resulted in higher degree of densification. The shrinkage rate (dL/dt) depended on the heating rate applied. For example, above 1000 °C the shrinkage rate for a 20 °C/min heating rate was several times greater than that for a heating rate of 3 °C/min, indicating that a film would be more likely to crack when sintered at a 20 °C/min heating rate.

An excellent sintering activity was found for the LSCF powder used in the present study. This was due to its high specific surface area and relatively homogeneous particle size distribution. The shrinkage rate reached a maximum at approximately 1200 °C. Considering a porosity of approximately 35 vol% was required for a cathode film, an appropriate sintering temperature for the LSCF films should fall in the range 800-1000 °C.

### 3.3.2.3 Modified sintering programs
If the cracks were generated by the constrained shrinkage of the films, then it was considered that the cracking problem might be avoided by optimizing the burn-out and sintering schedule based on the thermal analyses described above. The sintering program was therefore modified as follows: for the organics removal step between room temperatureand350 °C, a very slow heating rate was employed(3 °C/min or 1 °C/min, holding for 2 hours at 350 °C); for the stable period between 350and800 °Ca medium heating rate was used(5 °C/min) and a holding time was applied at 800 °C(4 hours) at the start of sintering to strengthen the particle network; between 800 and the target maximum sintering temperature a much lower heating rate was used(1 °C/min)and a much longer holding time (48 hours) was used at the top temperature; finally cooling to room temperature was done at 5 °C/min. For the particular case of a top temperature of 1000 °C, the sintering program is plotted in Fig.8, compared with the initial sintering schedule.

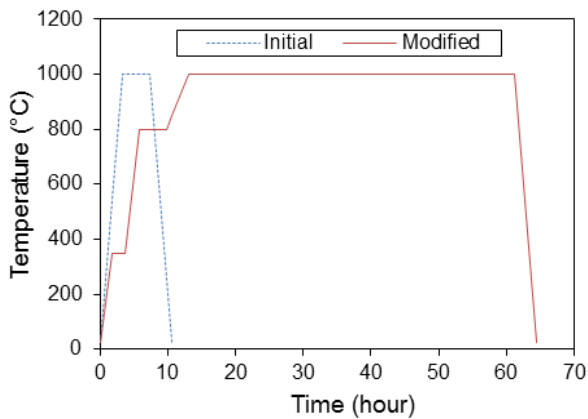

*Fig. 8. Comparison of the initial and modified sintering programs.*

The micrograph of the sintered film using the modified sintering program is shown in Fig.9. Unfortunately this shows little improvement of surface quality compared to that using the initial sintering program. There are still many cracks up to 1.5 μm wide and asperities at least 10 μm in diameter showing little effectiveness of the modified sintering conditions.

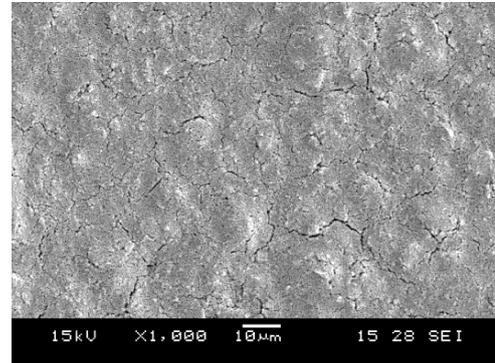

*Fig. 9. Micrograph of the surface features of a film sintered with the modified sintering program.*

Other sintering programs were also explored to see if they could prevent the cracking. Fig. 10 are the top surface micrographs of LSCF films sintered using these alternative sintering programs: 1) Sintering temperature 900 °C, holding for 4h, extremely slow heating rate of 5°C/h between 800-900°C,heating and cooling rate 5 °C/min for other segments; 2) Sintering temperature 800 °C, holding for 4h, extremely slow heating rate 3 °C/h between 700-800 °C, heating and cooling rate 5 °C/min for other segments; 3) Sintering temperature: 800C, holding for 20h, heating and cooling rate 6 °C/h.
Extensive cracks and asperitiescould still be found in all sintered films in the above experiments, indicating that the prevention of cracking was not achieved by lowering sintering temperatures or slowing down heating/cooling rates. This also implies that cracking in the films does not originate in the sintering process. However, the results do showthat a lower sintering temperature results in a smaller crack opening width. Fig. 11 shows the cracks in LSCF films after sintering at 1100 and 1200 °C. Compared with the film sintered at 1000 °C, the crack opening width increased to 6 and 10 μm, respectively, resulting in islands of approximately 30 μm diameter on the surface.
Our recent study [16] of the formation in 8 mol% yttria stabilized zirconia (8YSZ) films showed that cracks already present in the dried films cannot be healed during constrained sintering. which suggested that cracks found after sintering are probably initiated during the drying stage.

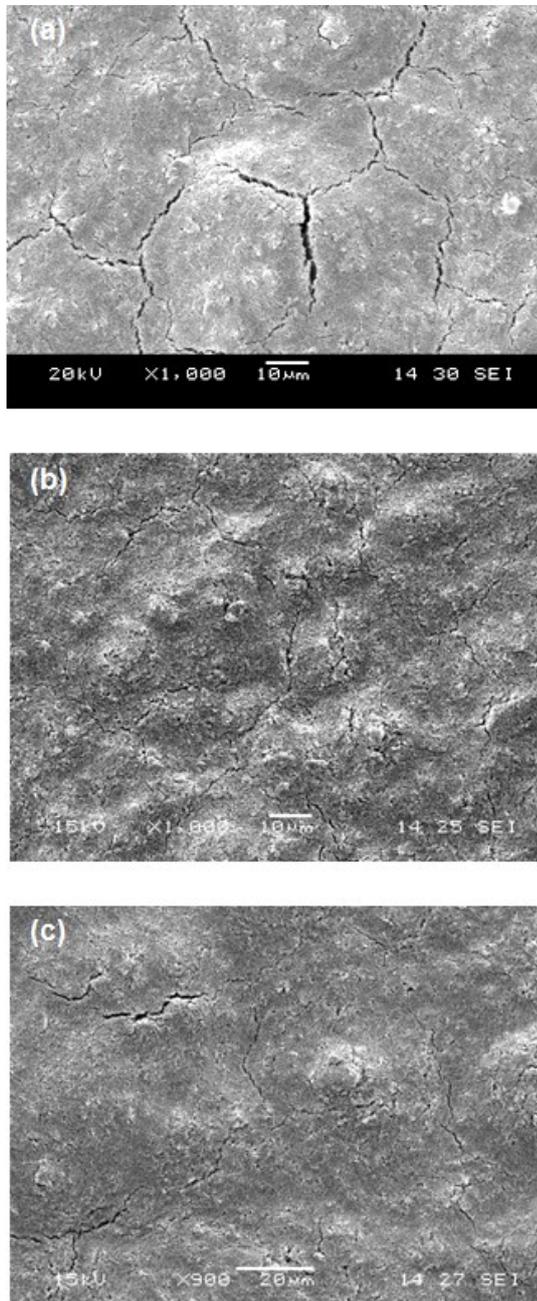

*Fig. 10. Micrographs of LSCF film top surfaces after different sintering schedules. (a): Sintering temperature 900 °C, holding for 4h, slow heating rate of 5 °C/h between 800-900 °C, heating and cooling rate 5 °C/min for other segments; (b) Sintering temperature 800 °C, holding for 4h, slow heating rate 3 °C/h between 700-800 °C, heating and cooling rate 5 °C/min for other segments; (c) Sintering temperature: 800C, holding for 20h, heating and cooling rate 6 °C/h*

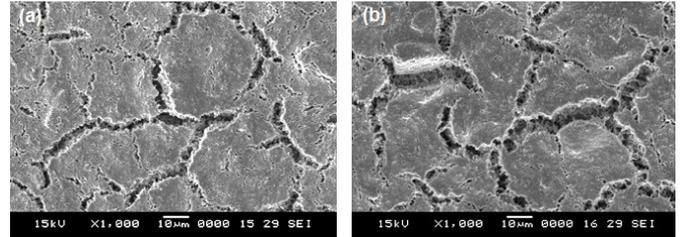

*Fig. 11. The surface cracks on LSCF films sintered at (a) 1100 °C and (b) 1200 °C.*

However, it can be very difficult to directly observe drying cracks because their opening width is often as small as the particle size. Fig.12 shows top surface micrographs of the as-dried LSCF films, in which it is difficult to identify any cracks with certainty, although typical asperities, probably due to inhomogeneous packing of particles, are evident at this early stage.

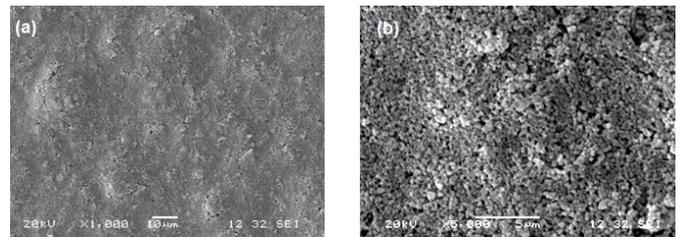

*Fig. 12. Micrographs of dried LSCF films screen printed on CGO substrates. Images (a) and (b) are shown at different magnifications.*

### 3.3.3 Film deposition methods

Considering that the non-uniform pressure distribution due to the wire mesh of the screen printer could be a reason for the presence of cracking and surface asperities, we investigated an alternative deposition method to deposit the ink, i.e. casting with a doctor blade. The CGO pellet was partially covered by a perimeter mask of a controlled thickness and a doctor blade coated with LSCF ink was pressed under constant pressure and swept smoothly across the pellet surface by hand. Drying and sintering programs, which were identical to the ones used for the screen-printed samples in the beginning of the study, were applied. The SEM micrographs of the sample surface after sintering are shown in Fig. 13. The surface is smoother and has finer cracks although the network remained visible. It is speculated that tape casting eased the homogeneous packing of the particles in the ink after deposition, so the presence of agglomerates was reduced, resulting in a flatter film surface, but the cracking was not prevented, although reduced in severity.

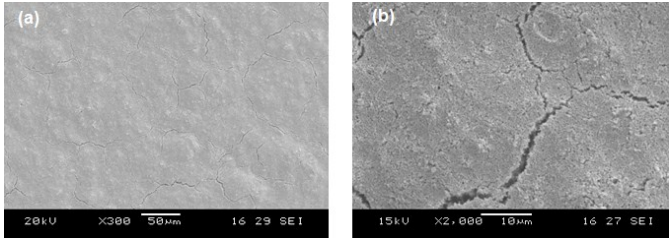

*Fig. 13. Micrographs of sintered LSCF films fabricated by tape casting. Images (a) and (b) are shown at different magnifications.*

### 3.3.4 Ink formulation

In the light of the above experiments, the wetting of the ink and the dynamic interaction of the particles in the ink, which were largely controlled by the ink viscosity, were thought to play an important role in the packing of the particles in the films after deposition and subsequent drying and sintering. Therefore, reformulated inks were prepared as described in Section 2 and deposited on CGO substrates by tape casting. The initial drying and 3-step sintering programs were applied to sinter the as-deposited films at 1000ºC. The surface and fracture cross-sectional morphology of these two films were examined under SEM, as shown in Fig. 14.

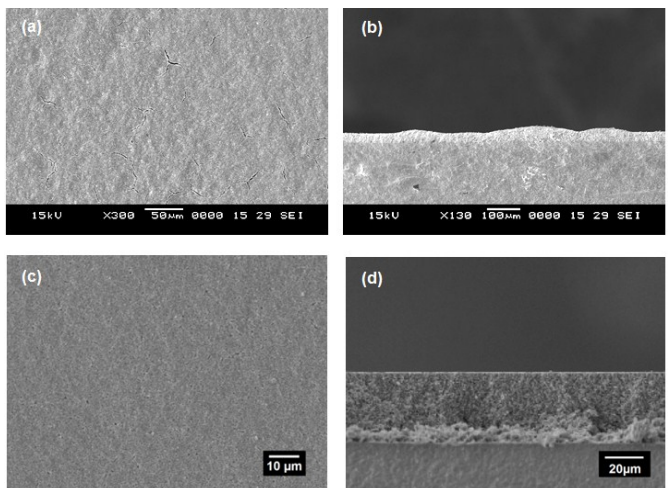

*Fig. 14. Micrographs of top surfaces and fracture cross-sections of as-sintered LSCF films made by tape casting reformulated inks: (a, b): 1:1 diluted ink; (c, d): 1:2 diluted ink.*

Compared with films (Fig. 1) made from the as-received ink, Fig. 16 shows a drastic decrease on the number of cracks and surface agglomerates. Some fine cracks and surface asperities could still be observed in the film made from the 1:1 diluted ink, whereas for the film made from 1:2 diluted ink, cracks are not seen and the film maintained a very flat surface. A more detailed observation using FIB cross sectioning and SEM (Fig. 15) shows that the cracks were completely eliminated in the latter film, showing that by using the 1:2 diluted ink acceptable non-defective films could be fabricated. It is worth noticing that compared with the microstructure shown in Figures 1 and 2 for the films made from the original ink, Figures 14 (c), (d) and 15 demonstrate a much more homogeneously interconnected microstructure for the films made from 1:2 diluted ink, and have benefitted from the lower ink viscosity.

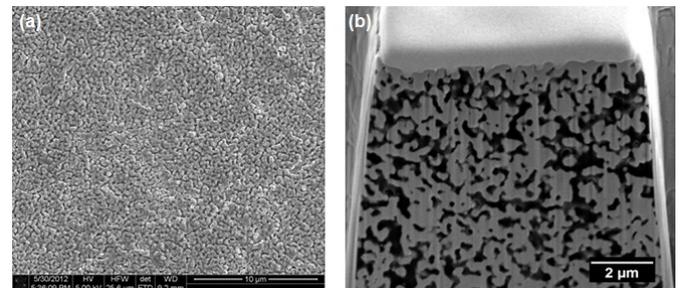

*Fig. 15. (a) High magnification surface and (b) FIB sliced cross-sectional micrographs of a film made from the 1:2 diluted ink after sintering at 1000 ºC. Note that the dark phase in (b) is epoxy resin impregnated prior to FIB slicing*

Nanoindentation tests were performed subsequently on this crack-free and flat film, the response curves and the resulting elastic modulus are plotted in Fig.16 (a) and (b), respectively. The results above show that the measurements for the non-defective film were significantly less scattered than for the defective one (Fig. 3), as the unloading response curves plotted in Fig. 16(a), from which the elastic modulus was determined, were highly consistent. Table 2 compares the properties of the as-receivedink and the diluted inks as well as the characteristics of the films fabricated from them.

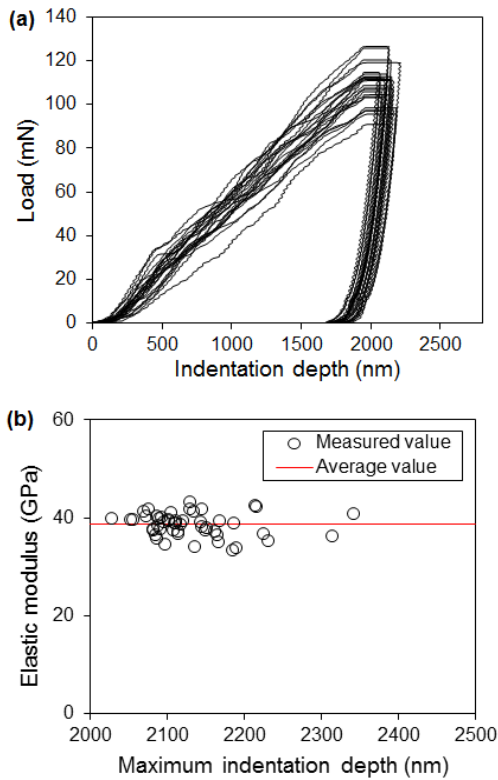

*Fig. 16. (a) Nanoindentation load vs. indentation depth curves and (b) the corresponding elastic modulus values calculated for the crack-free and flat film made from 1:2 diluted ink.*

*Table 2 Comparison between properties of the as-received and diluted inks and their corresponding films sintered at 1000 °C.*

| Properties | Original ink | 1:1 ink | 1:2 ink |
|---|---|---|---|
| Viscosity (*Pa.s*) | 41.51 | 12.36 | 7.19 |
| Corresponding films fabricated | | | |
| Average surface roughness $R_a$ (*μm*) | 1.82 | 0.89 | 0.20 |
| Elastic modulus (*GPa*) | 33.7±21.3 | 35.4±10.8 | 38.6±2.4 |

Lower ink viscosities resulted in remarkable decreaseof surface roughness andstandard deviation of elastic modulus, implying a significantly reducednumber of surface defects in the corresponding films. As can be seen in Table 2, the surface roughness for the films made from 1:2 ink was reduced to merely 10% that of films made from the as-receivedink.This is much smaller than the indentation depth, resulting in a steep reductionin the relative standard deviationofmeasured elastic modulus from 63.2% to 6.2%.

The experiments above revealed that rather than shrinkage during sintering, or differential contraction during cooling as proposed in many studies[11, 17], the more critical factor for obtaining crack-free and flat films in the current study was the ability of the ink to be self-leveling in the early wet state. Cracking is most likely to initiate at the drying stage if the particles are prevented from packing more effectively as the liquid content was removed. Lowering ink viscosity, by adding more terpineol solvent in this case, could effectively keep the ceramic particles in a stable suspension in the ink and flocculated agglomerates could be avoided. It also promoted the dispersion of the particles in a more homogeneous way as revealed by Maiti and Rajender [18]. As a result, cracking and asperity formation could be reduced to an acceptable level, or even be completely avoided. Consistent and reliable nanoindentation measurements could thus be conducted with higher reproducibility.

## 3. Conclusions

In summary, we have shown that sintered porous thin LSCF SOFC cathode films fabricated using a typical screen-printing ink were full of cracks and surface asperities, which caused extreme inconsistency and errors in attempts to measure film elastic modulus by nanoindentation. Various processing parameters were investigated in order to eliminate these defects. Different film deposition methods and thermal treatments showed little effect on the prevention of the defect formation. Furthermore the cracks were shown not to arise from thermal expansion mismatch between film and substrate and it was concluded that the defects initiated during drying of the ink and then made more severe by the sintering process. Thus, sintered porous thin LSCF cathode films were successfully fabricated without any crack or surface asperities by using a much less viscous ink. Results suggested that adequate amount of terpineol solvent added in the commercial ink eased the wetting and self-leveling of the ink upon deposition and thus a more homogeneous packing of LSCF particles in the films was achieved. As a result there was more homogeneous shrinkage in the film during the thermal treatments followed and lower local stresses in the films. With the non-defective films, consistent and reliable elastic modulus values of the films could then be obtained using nanoindentation.

## Acknowledgement

This research was carried out as part of the UK Supergen consortium project on "Fuel Cells: Powering a Greener Future". The Energy Programme is an RCUK cross-council initiative led by EPSRC and contributed to by ESRC, NERC, BBSRC and STFC. Z. Chen is especially grateful to the Chinese Government and Imperial College for financial support in the form of scholarships. Thanks are additionally due to Dr. Vineet Bhakhri for assistance with nanoindentation.